\RequirePackage{ifpdf}
\documentclass[paper]{JHEP3}
\usepackage{cite}
\usepackage{epsfig}
\usepackage{blkarray}
\usepackage{lscape}
\usepackage{booktabs,multirow,tabularx}

\newcommand{\matindex}[1]{\mbox{\scriptsize#1}}
\def\mymati#1{\matindex{${#1}$}}

\def\beq{\begin{equation}}
\def\beqn{\begin{eqnarray}}
\def\eeq{\end{equation}}
\def\eeqn{\end{eqnarray}}
\def\abs#1{\left|#1\right|}
\def\Mm#1#2{\mu_{#1}^{(#2)}}

\newcommand\sss{\scriptscriptstyle}
\newcommand\pt{p_{\sss T}}
\newcommand\kt{k_{\sss T}}
\newcommand\Ht{H_{\sss T}}
\newcommand\as{\alpha_{\sss S}}
\newcommand\mtop{m_{t}}
\def\GeV{\, \rm GeV}
\def\mtoppd{\mtop^{\rm pd}}
\def\mtopa{\mtop^{(\alpha)}}
\def\mtopb{\mtop^{(\beta)}}
\def\bmtopa{\overline{m}_{t}^{(\alpha)}}
\newcommand\bt{\bar{t}}

\newcommand\aNLO{{\sc\small MadGraph5\_aMC@NLO}}
\newcommand\HWs{{\sc\small HERWIG6}}
\newcommand\Madspin{{\sc\small MadSpin}}
\newcommand\kin{{\cal K}}
\newcommand\plus{\oplus}
\newcommand\muF{\mu_{\sss F}}
\newcommand\muR{\mu_{\sss R}}
\newcommand\xiF{\xi_{\sss F}}
\newcommand\xiR{\xi_{\sss R}}

\preprint{
 CERN-PH-TH-2014-117\\
 Cavendish-HEP-14/05
 }

\title{Determination of the top quark mass from leptonic observables}

\author{Stefano~Frixione$^a$ and Alexander~Mitov$^b$\\
 $^a$ PH Department, TH Unit, CERN, CH-1211 Geneva 23, Switzerland\\
 $^b$ Cavendish Laboratory, University of Cambridge, Cambridge CB3 0HE, UK
}

\abstract{
We present a procedure for the determination of the mass of the 
top quark at the LHC based on leptonic observables in
dilepton $t\bar{t}$ events. Our approach utilises the shapes of 
kinematic distributions through their few lowest Mellin moments; 
it is notable for its minimal sensitivity to the modelling of 
long-distance effects, for not requiring the reconstruction of 
top quarks, and for having a competitive precision, with theory errors 
on the extracted top mass of the order of 0.8~GeV. A novel aspect
of our work is the study of theoretical biases that might influence
in a dramatic way the determination of the top mass, and which are
potentially relevant to all template-based methods. We propose a
comprehensive strategy that helps minimise the impact of such biases,
and leads to a reliable top mass extraction at hadron colliders.
}

\begin{document}

\section{Introduction\label{sec:intro}}

The current world average of the top quark mass~\cite{ATLAS:2014wva}
\begin{equation}
\mtop = 173.34 \pm 0.76 \GeV ~~ [{\rm World\, Average}]
\label{eq:mt-value}
\end{equation}
implies that $\mtop$ is known with a precision better than $0.5 \%$. 
Such an accuracy is perfectly adequate for present collider-physics 
applications~\cite{Juste:2013dsa} including, notably, the global electroweak 
(EW) fits~\cite{Baak:2012kk}, which are saturated by the uncertainty on 
the $W$-boson mass, and not by that on $\mtop$. Still, the accurate 
determination of the top quark mass at hadron colliders remains a 
subject of much activity and debate.

Two separate developments have been the main drivers behind the above mentioned
activity: the outsize role played by the top quark mass in determining the
stability of the electroweak vacuum (both in the Standard Model 
(SM)~\cite{Bezrukov:2012sa,Degrassi:2012ry,Buttazzo:2013uya} and 
beyond~\cite{Bezrukov:2007ep}), and the recognition that the extraction 
of $\mtop$ at hadron colliders involves significant theoretical challenges, 
that might conceivably affect its value at the level of ${\cal O}(1 \GeV)$ 
(see ref.~\cite{Juste:2013dsa} for detailed discussion).

The bottom-up extrapolation of EW-scale physics, based on
eq.~(\ref{eq:mt-value}), implies either that the EW vacuum becomes unstable
below the Planck scale, or that the result of eq.~(\ref{eq:mt-value}) deviates
from the value needed for the stability of the SM EW vacuum up to the Planck
scale by about two to four sigma's~\cite{Buttazzo:2013uya,Kobakhidze:2014xda}. 
If confirmed, such a conclusion might indirectly imply the existence of Beyond
the SM (BSM) physics somewhere below the Planck scale. Given the non-observation
of BSM signals so far, it would be hard to overstate the importance of 
this implication. We stress that these facts are mainly driven by the 
$\mtop$ value of eq.~(\ref{eq:mt-value}), and this because of the large
parametric dependence of the stability condition on the top quark mass.

At this point one might wonder about the need for revisiting the subject of
$\mtop$ determination, given the quite high precision of the top mass 
of eq.~(\ref{eq:mt-value}). To this end let us remind the reader that there 
are a number of high-precision measurements that marginally agree with the 
current world average. Examples are the very recent CMS~\cite{CMS-mass-2014} 
and D0~\cite{Abazov:2014dpa} measurements:
\begin{eqnarray}
&& \mtop = 172.04 \pm 0.77 \GeV ~~ [{\rm CMS\, Collaboration}]\, , \nonumber\\
&& \mtop = 174.98 \pm 0.76 \GeV ~~ [{\rm D0\, Collaboration}]\, . 
\label{eq:mt-value-CDF-D0}
\end{eqnarray}
The above measurements have the same uncertainty as the combination in
eq.~(\ref{eq:mt-value}), but notably different central values\footnote{The
measurements in eq.~(\ref{eq:mt-value-CDF-D0}) agree with the world average 
of eq.~(\ref{eq:mt-value}) at approximately $2\sigma$.}.
In particular, the CMS measurement~\cite{CMS-mass-2014} is consistent with the
SM EW vacuum being stable up to the Planck scale, while the D0 
measurement~\cite{Abazov:2014dpa} implies a rather unstable SM EW vacuum. 
Therefore, the spread in the available $\mtop$ measurements alone warrants a
closer inspection of the determination of the top quark mass. As we shall
detail in the following, there are also strong theoretical reasons that
motivate further studies of the extraction of this parameter from hadron
collider data.

The determination of the top quark mass is as much dependent on theoretical
assumptions as it is on measurements. The reason is that the top quark mass 
is not an observable and thus cannot be measured 
directly\footnote{For this reason we do not speak of top mass 
{\em measurements} but of top mass {\em determinations} or 
{\em extractions}.}: it is a theoretical concept, and its value 
is extracted from data in collider events that feature top quarks. 
Such an extraction depends on the definition of the mass (pole mass,
running mass, and so forth), on the observables chosen, and on the various
approximations made when computing those observables. Since measurements 
are insensitive to theory assumptions\footnote{Strictly speaking, this is 
never the case. For example, corrections for detector effects do depend on
theory assumptions. In the first approximation, one can ignore these
data-theory correlations.}, any modification in the theoretical 
modelling will result in a different value of the extracted top mass. 
If everything is consistent, i.e.~if the estimated uncertainty is a realistic
representation of the true uncertainty, then the differences in the returned
values should fall within the corresponding theory errors. In reality, this
may not be the case due to the presence of biases, whose very existence
might be difficult to establish. With this important subtlety in mind, 
one of the main aspects of the present work is to devise a structured
approach towards the identification of such hidden biases.

A significant number of techniques for the determination of the top mass 
exist or are under study; see ref.~\cite{Juste:2013dsa} for a recent 
in-depth overview. Such techniques may be organised into two classes,
whose definitions cannot be given in a rigorous way, but which are
nevertheless based on clearly distinct physical principles.
The first class includes all those approaches that use, in some
form, the fact that the top is a particle that decays: the knowledge
of the decay products (i.e.~their identities and kinematic configurations)
is then exploited to reconstruct some quantity which is directly related
to the top, and thus bears information on its mass. The crucial characteristic
is that, by emphasising the role of the decay, one factors out the
details of the process in which the primary top(s) is(are) produced,
so that the details of the production mechanism become irrelevant.
The ideal (i.e.~not realistic) procedure which belongs to this class 
is the one where the top virtuality is reconstructed exactly by measuring 
the invariant mass of its decay products, thus scanning its lineshape. 
In the approaches that belong to the
second class the role of the top as a mother particle must not matter;
the only important thing is that some observable(s) of a top-mediated
process depend in a significant way on $\mtop$, so that their measurements
can be mathematically inverted (using suitable theoretical predictions)
to return the top mass. We stress that the fact that the observables 
mentioned above are most likely constructed by using the top decay products 
is not relevant. The only important thing is that they depend on the
top quark mass, a feature that might be possessed by other quantities
as well (for example, the primary QCD radiation in the production process).

The approaches that belong to the first class are often perceived to
be affected by smaller theoretical systematics than those of the second
class, because by their very definition one assumes that many sources
of uncertainties, such as PDF dependence, absence of higher-order
perturbative corrections, and new-physics contributions, will drop out,
being mostly associated with the production mechanism. Unfortunately,
this is not really the case. Firstly, some of these sources might be
relevant to decays as well. Secondly, different kind of uncertainties
could become important: a good example is the so-called $J/\Psi$ 
method~\cite{Kharchilava:1999yj} which, although experimentally very 
clean and theoretically well defined, is hampered by its sensitivity 
to the non-perturbative $b$-fragmentation. Thirdly, in these approaches
one must start by {\em defining} what one means by ``top'', which
introduces some auxiliary (if only intermediate) concept in the procedure,
and renders it difficult to assign a proper theoretical error to it.
Note that this necessity goes beyond what one must do in order to
reconstruct the top quark experimentally, and is purely theoretical.

The bottom line is that, regardless of which class an $\mtop$-extraction
technique belongs to, some amount of theoretical modelling will be
involved. In this paper, we follow an approach of the second class;
we believe that not having to define the top as a final-state object
is a virtue that more than compensates a larger dependence on the 
production process.

Another important motivation behind the procedure we are proposing
is the use of observables that can be both reliably predicted within 
the SM, and cleanly measured. Thus, we employ kinematic distributions 
of leptons in dilepton $t\bt$ events; more precisely, we are interested 
in their shapes. Furthermore, we find that the information on the top mass 
that such shapes encode can be very effectively provided by the Mellin 
moments of the corresponding distributions, and it is such moments that
will play a central role in our method. 
Our goal is the determination of $\mtop$ with competitive precision,
supplemented by a detailed study of the various sources of theoretical
systematics. Apart from not having to rely, directly or indirectly, on
the reconstruction of top quarks, our approach has minimal 
sensitivity to the modelling of both perturbative and non-perturbative QCD 
effects\footnote{The emphasis is on ``modelling'' here: we point out
that in parton shower Monte Carlos several choices can be made (e.g.~those
of the so-called shower variables) that affect the perturbative part of the 
simulation, which are all compatible with the underlying perturbative 
description.}. We believe that the latter property is one of the chief 
advantages of the method we are pursuing.

In this paper we shall be working with the top quark pole mass, and 
shall not consider alternative mass definitions. Our viewpoint is that the
intrinsic differences between any two of these definitions (renormalon-related
effects are a good example) are largely below the present level of
uncertainties, and therefore we do not see them as a reason for concern at
present. A fuller discussion can be found in ref.~\cite{Juste:2013dsa}. 

We shall conclude that, with the procedure we employ, the extraction of
the top pole mass can be achieved with a theoretical error of about $0.8\GeV$,
and possibly smaller. While a significant number of $t\bt$ dilepton events
have been recorded during Run I of the LHC, no measurements are published 
of the Mellin moments that would allow us to apply our procedure to
real data. We thus hope that this paper will encourage the LHC
experimental collaborations to measure directly such moments, so that 
the present analysis could be repeated, and its results compared with
those of eqs.~(\ref{eq:mt-value}) and~(\ref{eq:mt-value-CDF-D0}).
Furthermore, we are hopeful that the reliability and small theoretical
systematics of the method proposed in this work will help shed light 
on the issue of the EW vacuum stability.

This paper is organised as follows: in sect.~\ref{sec:method} we introduce
our method in detail and define, in particular, its associated theoretical
errors (sect.~\ref{sec:extract}) and biases that may affect it 
(sect.~\ref{sec:bias}). Our results are presented in sect.~\ref{sec:results}:
those with the highest theoretical accuracy in sect.~\ref{sec:res},
with discussions on the effects due to parton showers, higher orders,
and spin correlations in sect.~\ref{sec:eff}, and explicit 
examples of theoretical biases in sect.~\ref{sec:resbias}. We give
our conclusions in sect.~\ref{sec:concl}. Some technical material
is reported in the appendices.

\section{The method}\label{sec:method}
Our goal is to study the determination of the top quark pole mass $\mtop$
from several differential distributions of leptons in dilepton $t\bt$ events:
\beq
pp \to t\bt +X\,,\;\;\;\;\;\;\;\;
t\to \ell^+\nu_\ell b\,,\;\;\;\;
\bt\to \ell^-\bar{\nu}_\ell \bar{b}\,,\;\;\;\;\;\;\;\;
\ell=e,\mu \,.
\label{eq:reaction}
\eeq
Each of the observables that we consider features the following
important properties:
\begin{itemize}
\item It does not require the reconstruction of the $t$ and/or $\bt$ quark;
indeed, we do not even need to speak of top quarks\footnote{We shall ignore 
backgrounds. In a more realistic analysis, some mild dependence on the 
definition of top quarks might enter through the subtraction of backgrounds.}.
\item It is almost completely inclusive in hadronic radiation:  the
only possible dependence on strongly-interacting final-state objects
is that due to selection cuts (on $b$-jets).
\item Owing to this inclusiveness, the observable is minimally sensitive 
to the modelling of long-distance effects. This feature increases the 
reliability of the theoretical predictions.
\end{itemize}
The set of observables considered in this paper and their labelling 
conventions are 
given in table~\ref{tab:obs}: $\pt(\ell^+)$ is the single-inclusive
transverse momentum of the positively-charged lepton; $\pt(\ell^+\ell^-)$
and $M(\ell^+\ell^-)$ are the transverse momentum and the invariant
mass, respectively, of the charged-lepton pair; finally, $E(\ell^+)+E(\ell^-)$
and $\pt(\ell^+)+\pt(\ell^-)$ are the scalar sums of the energies and
transverse momenta of the two charged leptons, respectively. We point out
that the latter two sums are computed event-by-event; in other words,
observables \#4 and \#5 are {\em not} constructed a-posteriori given
the single-inclusive energy and transverse momentum distributions of
the leptons.
\begin{table}
\begin{center}
\begin{tabular}{c|c}
Label & Observable
\\\hline
1 & $\pt(\ell^+)$ \\
2 & $\pt(\ell^+\ell^-)$ \\
3 & $M(\ell^+\ell^-)$ \\
4 & $E(\ell^+)+E(\ell^-)$ \\
5 & $\pt(\ell^+)+\pt(\ell^-)$\\ 
\end{tabular}
\end{center}
\caption{\label{tab:obs}
The set of observables used in this paper, and their labelling conventions.
}
\end{table}

The extraction of the top quark mass utilises the sensitivity of the
{\em shapes} of kinematic distributions to the value of $\mtop$. 
It is cumbersome to work directly with differential distributions. Instead, 
we utilise their lower Mellin moments, whose precise definition is given
in sect.~\ref{sec:definemoments}. The idea of the method proposed in 
this paper is to predict the $\mtop$ dependence of the moments, and then to
extract the value of $\mtop$ by comparing the predicted and measured values of
those moments. The procedure is detailed in sect.~\ref{sec:extract}.

The use of moments for the extraction of the top mass has been 
suggested previously in the context of the so-called $J/\Psi$ 
method~\cite{Kharchilava:1999yj}, or in connection with 
variables supposed to minimise the dependence on the jet-energy
scale~\cite{Hill:2005zy,Garberson:2008te}. To our knowledge, the
most up-to-date theoretical treatment of this technique is in
ref.~\cite{Biswas:2010sa}. All these papers consider only the
first moment (of various distributions); in the case 
of $\mtop$ extraction from different
observables, the results are either not combined~\cite{Biswas:2010sa},
or limited to two observables~\cite{Garberson:2008te}. These choices
may lead to issues, as we shall discuss in sects.~\ref{sec:bias},
\ref{sec:eff}, and~\ref{sec:resbias}. In the case of the dilepton 
channel, ref.~\cite{Biswas:2010sa} also employs one of the observables
considered in this paper ($E(\ell^+)+E(\ell^-)$); owing to the different
choices made for cuts, jet algorithm, collider energy, and PDFs,
we have refrained from making a direct comparison with those results. 
We also point out
that in ref.~\cite{Biswas:2010sa} the simultaneous variation of the
factorisation and renormalisation scales has been adopted, which
leads to smaller scale uncertainties than those we find in this
paper (where the two scales are varied independently, see 
sect.~\ref{sec:results}). 

Finally, we remark that other discrete 
parameters of kinematic distributions, such as medians and maxima, might
also be used for a top mass extraction. We have chosen to work with moments
because of the ease of their calculation, and of the fact that the results
they give can be systematically improved by including those of
increasingly higher rank. For other previous theoretical approaches
whose philosophy differs, in one or more aspects, w.r.t.~the one
adopted here, see e.g.~refs.~\cite{Frederix:2007gi,Kawabata:2011gz,
Alioli:2013mxa,Kawabataa:2014osa}.

\subsection{Definition of moments\label{sec:definemoments}}

We denote by $\sigma$ and $d\sigma$ the total and fully-differential
$t\bt$ cross sections respectively (possibly within cuts), so that:
\beq
\sigma=\int\! d\sigma\,,
\eeq
where the integral in understood over all degrees of freedom.
Given an observable $O$ (e.g.~one of those in 
table~\ref{tab:obs}), its normalised  moments are defined 
as follows:
\beq
\Mm{O}{i}=\frac{1}{\sigma}\int\! d\sigma\,O^{\,i}\,,
\label{mmdef}
\eeq
for any non-negative integer $i$. In this way, one has:
\beq
\Mm{O}{0}=1\,,\;\;\;\;\;\;\;\;
\Mm{O}{1}=\langle O\rangle\,,\;\;\;\;\;\;\;\;
\Mm{O}{2}=\langle O^2\rangle=\sigma_O^2+\left(\Mm{O}{1}\right)^2\,,
\eeq
and so forth. Note that, when selection cuts are
applied (see eq.~(\ref{eq:cuts})) in the calculation of moments,
they are applied exactly in the same manner in the denominator and 
in the numerator of eq.~(\ref{mmdef}).

A short technical remark: the numerator of eq.~(\ref{mmdef}) is usually
derived from the result relevant to the differential distribution
\mbox{$d\sigma/dO$}. On the other hand, it could also be computed
directly (i.e., without using \mbox{$d\sigma/dO$}), which is a procedure
affected by smaller uncertainties, as we explain in appendix~\ref{sec:comp}.
The important thing to point out here is that such a direct calculation
has a fully analogous experimental counterpart: Mellin moments can be
measured directly, which is the procedure we recommend. See 
appendix~\ref{sec:comp} for more details.

\subsection{Extraction of the top quark mass and its 
uncertainties\label{sec:extract}}

\begin{figure}[h]
 \begin{center}
 \epsfig{file=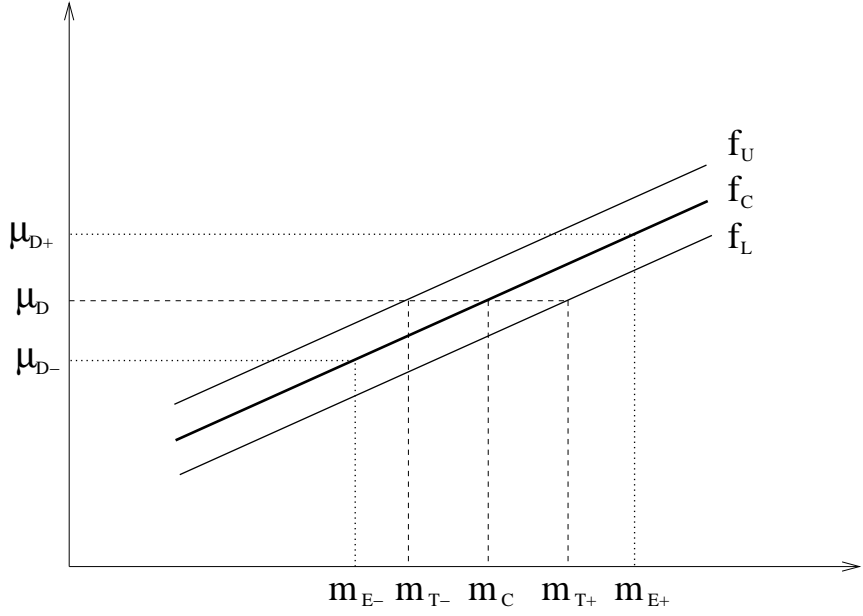, width=0.75\textwidth}
 \end{center}
\caption{Graphic representation of the method used in this paper
to extract the top mass from the first moment of any given observable.
The case of higher moment is identical, except for the fact that the
$x$ axis is associated with $\mtop^i$.}
\label{fig:extract}
\end{figure}

The method for extracting $\mtop$ from the first moment of any one 
of the observables of table~\ref{tab:obs} is given schematically in
fig.~\ref{fig:extract}. 
The $x$ and $y$ axes of fig.~\ref{fig:extract} are associated with the top 
pole mass $\mtop$, and the first moment of $O$, $\Mm{O}{1}$, respectively. 
The three lines $f_C$, $f_U$, and $f_L$ represent the central, upper, and 
lower theoretical predictions for $\Mm{O}{1}(\mtop)$. The case of moments
higher than the first, $\Mm{O}{i}(\mtop)$, $i>1$, is identical, except
for the fact that the $x$ axis of fig.~\ref{fig:extract} is associated 
with $\mtop^i$ (see eq.~(\ref{eq:res-equation})). 
We have assumed that $\Mm{O}{i}$ increases with $\mtop$, 
which is indeed the case in the SM and for the observables considered
here; the formulae given below can however be trivially extended to the
case of $\Mm{O}{i}$ decreasing with $\mtop$.
As fig.~\ref{fig:extract} suggests, the functions $f_{C,U,L}$ are
linear in $\mtop^i$; although in general they need not be so, we have 
found that straight lines are perfectly adequate to our purposes.
We explain how such functions are computed in sect.~\ref{sec:thsetups}.

Given the data\footnote{Despite the large number of $t\bt$ dilepton 
events accumulated so far at the LHC, no measurement of these moments 
is available at present.}
\beq
\mu_D {}_{-\Delta_\mu^-}^{+\Delta_\mu^+}\,,
\label{mudata}
\eeq
with
\beq
\Delta_\mu^-=\mu_D-\mu_{D-}\,,\;\;\;\;\;\;\;\;
\Delta_\mu^+=\mu_{D+}-\mu_D\,,
\eeq
the extracted top mass will be (see fig.~\ref{fig:extract}):
\beq
\mtop=m_C {}_{-\Delta_{mT}^-}^{+\Delta_{mT}^+}
{}_{-\Delta_{mE}^-}^{+\Delta_{mE}^+}\,.
\label{mextr}
\eeq
We define the central value and theoretical uncertainties associated with 
such an extraction as follows:
\beq
\Delta_{mT}^-=m_C-m_{T-}\,,\;\;\;\;\;\;\;\;
\Delta_{mT}^+=m_{T+}-m_C\,,
\label{therr}
\eeq
with
\beq
m_C=f_C^{-1}\left(\mu_D\right)\,,\;\;\;\;\;\;\;\;
m_{T-}=f_U^{-1}\left(\mu_D\right)\,,\;\;\;\;\;\;\;\;
m_{T+}=f_L^{-1}\left(\mu_D\right)\,.
\label{inv}
\eeq
Since the functions $f_{C,U,L}$ are linear in $\mtop^i$, their inversion
is trivial; however, we point out that eq.~(\ref{inv}) remains
valid regardless of the particular (monotonic and increasing) forms 
of $f_{C,U,L}$. While the quantities introduced in 
eq.~(\ref{therr}) are the theory errors that affect the top-mass 
extraction from any given observable and moment, there might be cases
in which they are inadequate to measure the actual difference between
the central value $m_C$ and the physical top mass. 
This happens in the presence of what we call theory {\em biases}, 
which we shall discuss at length in sect.~\ref{sec:bias}.

In keeping with fig.~\ref{fig:extract}, we define the experimental errors as:
\beq
\Delta_{mE}^-=m_C-m_{E-}\,,\;\;\;\;\;\;\;\;
\Delta_{mE}^+=m_{E+}-m_C\,,
\eeq
with
\beq
m_{E-}=f_C^{-1}\left(\mu_{D-}\right)\,,\;\;\;\;\;\;\;\;
m_{E+}=f_C^{-1}\left(\mu_{D+}\right)\,.
\eeq
It is easy to convince oneself that the more conservative choice:
\beq
m_{E-}=f_U^{-1}\left(\mu_{D-}\right)\,,\;\;\;\;\;\;\;\;
m_{E+}=f_L^{-1}\left(\mu_{D+}\right)\,,
\eeq
is not correct, since it leads to non-zero uncertainties also 
in the case of null experimental errors. 
In this paper, we shall not consider the experimental uncertainties
any longer, and shall be concerned only with the theoretical ones.
We point out that the size of these depend on two factors:
the uncertainty on the theoretical predictions for $\Mm{O}{i}$,
which is \mbox{$f_U(\mtop)-f_C(\mtop)$} or \mbox{$f_C(\mtop)-f_L(\mtop)$}
at a given $\mtop$, and the slope of $f_C(\mtop)$: the steeper the latter, 
the smaller the errors on the extracted values of $\mtop$.

\subsection{Theory biases\label{sec:bias}}
In this section we address the question whether there might be some biases
in the method outlined in sect.~\ref{sec:extract}, that would prevent the 
errors defined in eq.~(\ref{therr}) from being a reliable representation 
of the true uncertainties underlying the $\mtop$ extraction.

It is not difficult to devise a scenario where the answer to the
previous question is positive. Let us suppose that $t\bt$ production,
as is seen in LHC detectors, proceeds through both the well-known SM
mechanisms, and the exchange of a hypothetical heavy non-SM resonance. 
The nature of the
latter is irrelevant here; what matters is the fact that, owing to
its being very massive, it will cause the $t$ and $\bt$, and hence
their decay products, to be slightly more boosted on average
than if only SM physics would exist. Thus, for example, the measured 
first moment of $\pt(\ell^+)$ (which is observable \#1 in 
table~\ref{tab:obs}) will have a larger value than what would
be measured if only the SM were present. Let us further suppose that
the theoretical predictions used to extract $\mtop$ with the procedure
of sect.~\ref{sec:extract} are the pure-SM ones: what has been said above
also implies that the functions $f_C$, $f_U$, and $f_L$ will have lower
values, for any given top mass, than their counterparts computed in the
BSM theory that corresponds to the measured data. Figure~\ref{fig:extract}
then leads immediately to the conclusion that the value of $\mtop$ extracted 
from BSM data using SM predictions will be larger than the ``true''
top mass value\footnote{In keeping with sect.~\ref{sec:extract},
we have used the fact that the first moment of $\pt(\ell^+)$ is a growing 
function of $\mtop$. Qualitatively, the conclusions drawn in this section
are independent of the slope of $f_{C,U,L}$.}: the difference
between the extracted and the true $\mtop$ is then a theoretical bias.
The crucial point is that the uncertainties associated with such an 
extraction will be essentially the same\footnote{This is because $f_U$ 
and $f_L$ are, to a very good extent, parallel to $f_C$.} as those one 
would obtain in the complete absence of BSM physics in the data: in other 
words, the errors of eq.~(\ref{therr}) would fail to capture the presence
of the existing theoretical bias.

The main lesson to be drawn from the previous example is the following.
Given only one observable and one of its Mellin moments, the extraction 
of the top mass according to eq.~(\ref{mextr}) will always be possible 
with ``small'' theoretical errors\footnote{In a reasonable range of 
$\mtop$. For too large or too small $\mtop$ values this statement is not 
necessarily true: the functions $f_{U,L,C}$ might simply be inadequate to 
describe correctly the dependence of the moments on the top mass.}, regardless 
of whether the theory employed gives a correct representation of the 
physics model embedded in Nature. This observation, however, implicitly
suggests a solution to the problem posed by theoretical biases.
In fact, while the above indeed applies to each individual 
observable-and-moment choice, if the theoretical description is
ultimately not compatible with Nature it is not likely that
two values of $\mtop$ extracted from two different observables will
be compatible with each other. Conversely, the probability that
the extracted values of $\mtop$ be all mutually compatible (in the 
case of an incorrect underlying theoretical model) will decrease with 
the number of observables and moments considered. 

Note that it is easier to establish the possible incompatibility of any 
two $\mtop$ results when their theoretical errors are small. Therefore, 
the property of the uncertainties of eq.~(\ref{therr}) of being insensitive 
to theoretical biases is actually a positive feature in this context,
and underlines the importance of accurate predictions.
The bottom line is that, in order for the presence of theoretical
biases to be clearly uncovered, it is of utmost importance to consider
as many observables and moments as is possible. The choice of the
set of table~\ref{tab:obs} reflects this view, but it is clear
than any further addition to it will be beneficial.

We conclude this section by making various further observations.
Firstly, it is not necessary to
have a BSM-vs-SM scenario for theoretical biases to appear: it is
sufficient that theory and data are not fully compatible. We shall
give several examples of this in sect.~\ref{sec:results}, all of them
within the SM. Secondly, although possibly biased, the $\mtop$ value
extracted in a single-observable-and-moment procedure is not ``wrong'':
it is, by construction, the result that, for the given data, will give
the best prediction with the assumed theoretical model. Therefore,
as long as one uses such $\mtop$ with that model and only for that 
observable, one is perfectly consistent. It is in the interpretation
of the results, however, that one must be careful, since the larger
the bias, the less the extracted top quark mass will have to do with the
fundamental parameter so important e.g.~for the stability of the vacuum.
This stresses again the fact that it is always recommended, for example
through the multi-observable approach advocated here, to determine
the presence of theoretical biases. Thirdly and finally, the relationship
between $\mtop$-extraction and biases is by no means specific to the
use of Mellin moments; it is common to all template-based methods.
If anything, Mellin moments just render the discussion 
particularly transparent.

\section{Results\label{sec:results}}
Our predictions are obtained by simulating $t\bt$ production in the SM,
by treating the top quarks as stable, and by decaying them afterwards.
We perform the calculations in the fully automated 
\aNLO\ framework~\cite{Alwall:2014hca},
where we can easily investigate the impact of the various approximations
that may be employed; in particular, we shall consider both LO and NLO
results, with or without their matching to parton showers, with
or without including spin-correlation effects. We have thus several
calculational scenarios, which we summarise in table~\ref{tab:calc}.
We shall refer to each of them interchangeably with either their
labels or their extended names, the latter chosen in agreement with 
ref.~\cite{Alwall:2014hca}. NLO fixed-order computations are based
on the FKS subtraction method~\cite{Frixione:1995ms,Frixione:1997np}.
NLO results are matched to parton showers 
according to the MC@NLO formalism~\cite{Frixione:2002ik};
throughout this paper, we have used \HWs~\cite{Corcella:2000bw,
Corcella:2002jc}. Spin-correlation effects in the computations 
matched to parton showers are accounted for with the method of 
ref.~\cite{Frixione:2007zp} through its implementation in 
\Madspin~\cite{Artoisenet:2012st} (shortened to MS henceforth), 
a package embedded in \aNLO. As far as fixed-order results are
concerned, only decay spin correlations (i.e.~those described
by the matrix elements relevant to $t\to \ell^+\nu_\ell b$) are 
taken into account, whence the ``No'' in the rightmost entry of
the last two rows of table~\ref{tab:calc}.
\begin{table}
\begin{center}
\begin{tabular}{c|c||ccc}
\multirow{2}{*}{Label} & Extended & \multirow{2}{*}{Accuracy} & Parton & Spin \\
                       & name     &       & shower & correlations \\
\hline
1 & NLO+PS+MS & NLO & Yes & Yes \\
2 & LO+PS+MS  & LO  & Yes & Yes \\
3 & NLO+PS    & NLO & Yes & No \\
4 & LO+PS     & LO  & Yes & No \\
5 & fNLO      & NLO & No  & No \\
6 & fLO       & LO  & No  & No \\
\end{tabular}
\end{center}
\caption{\label{tab:calc}
Calculational scenarios considered in this paper. The rightmost column
reports the inclusion of {\em production} spin correlations; decay
spin correlations are included in all cases.
}
\end{table}

We have used a five-light-flavour scheme, and
the MSTW2008 (68\% CL) PDF sets~\cite{Martin:2009iq}
and their associated errors, at the LO or the NLO depending on the 
perturbative accuracy of the various scenarios reported in 
table~\ref{tab:calc}. We have included both PDF and
scale uncertainties in our predictions; both have been computed with
the reweighting method of ref.~\cite{Frederix:2011ss}. As far as
the latter uncertainties are concerned, they have been obtained with 
an independent variation of the renormalisation and factorisations scales, 
subject to the constraints
\beq
0.5 \leq \xiF\,,\;\xiR\,,\;\xiF/\xiR \leq 2 \, ,
\label{eq:scalevar}
\eeq
where
\beq
\muF=\xiF\hat\mu\,,\;\;\;\;\;\;\;\;
\muR=\xiR\hat\mu\,,
\label{muFmuR}
\eeq
and $\hat\mu$ is a reference scale; the default values or central scale
choices correspond to $\xiF=\xiR=1$. We point out that eq.~(\ref{eq:scalevar}) 
is a conservative scale variation (as was done e.g.~in
ref.~\cite{Cacciari:2008zb}, and as opposed to setting the two scales
equal to a common value), which estimates well the missing 
higher-order corrections to the total $t\bar t$ cross section at the 
NNLO~\cite{Czakon:2013goa,Czakon:2013xaa}. We have considered
three different functional forms for the reference scale $\hat\mu$
in eq.~(\ref{muFmuR}):
\beqn
&& \hat\mu^{(1)} = {1\over 2} \sum_{i} m_{{\sss T},i}\,,
\;\;\;\;\;\;i\in\{t,\bt\}\,,
\label{eq:scalesMtMT}
\\
&& \hat\mu^{(2)} = {1\over 2} \sum_{i} m_{{\sss T},i}\,,
\;\;\;\;\;\;i\in~{\rm final}~{\rm state}\,,
\label{eq:scalesHT}
\\
&& \hat\mu^{(3)} = \mtop\,,
\label{eq:scalesmt}
\eeqn
with the transverse masses $m_{{\sss T},i}=\sqrt{p_{{\sss T},i}^2+m_i^2}$.
We point out that, since in our calculations the top quarks are treated
as stable particles at the level of hard matrix elements, the difference 
between eq.~(\ref{eq:scalesMtMT}) and~(\ref{eq:scalesHT}) is the contribution 
to the latter of the transverse momentum of the massless parton which is 
possibly present in the final state (owing to real-emission corrections);
the scale of eq.~(\ref{eq:scalesHT}) is nothing but $\Ht/2$.

Our simulations are carried out at the 8 TeV LHC. Since we only consider
the process of eq.~(\ref{eq:reaction}), i.e.~top-pair production without
any background contamination, all of our events are $t\bt$ ones
by construction. On the other hand, in order to perform a more realistic
analysis, we also impose the following event selection: on top of
having two oppositely-charged leptons (electrons and/or muons),
events are required to contain at least two $b$-flavored jets, with
jets defined according to the anti-$\kt$ algorithm~\cite{Cacciari:2008gp} 
with $R=0.5$, as implemented in {\sc\small FastJet}~\cite{Cacciari:2011ma}. 
The events so selected are then subject to the following cuts:
\beqn
&& \abs{\eta(\ell^\pm)}\leq 2.4\,,\;\;\;\;\;\;\;\;
\pt(\ell^\pm) \geq 20~{\rm GeV}\,,
\nonumber\\
&& \abs{\eta(J_b)}\leq 2.4\,,\;\;\;\;\;\;\;\;\;
\pt(J_b) \geq 30~{\rm GeV}\,.
\label{eq:cuts}
\eeqn
If more than two $b$-jets are present, the cuts above are imposed on the
two hardest ones. In order to simplify our analysis, $b$-hadrons
have been set stable in \HWs, so that the vast majority of the
events just contain the two charged leptons arising from top decays.
In addition to the cuts of eq.~(\ref{eq:cuts}), we have also checked
the effects of imposing lepton-jet isolation cuts: these being negligible,
we shall not consider them any further in this paper.

\subsection{Calculation of the moments and
of the functions $f_{C,U,L}(\mtop)$\label{sec:thsetups}}
With the settings described above, we have simulated $t\bt$ production in all 
of the six calculational scenarios of table~\ref{tab:calc}; in the case of
NLO+PS+MS (which we believe to give the best description of SM physics,
and is thus treated as our reference computation), results have been 
obtained with all of the three scales choices of 
eqs.~(\ref{eq:scalesMtMT})--(\ref{eq:scalesmt}), while in all the other 
cases only the scale of eq.~(\ref{eq:scalesMtMT}) has been considered.

Each of these calculations has been performed eleven times, 
once for each value of the top quark mass chosen in the discrete set:
\beq 
\mtop=(168, 169, \dots, 178)~{\rm GeV}\,.
\label{eq:mtopset}
\eeq
In each of these runs, we have computed the first four Mellin moments
for all the observables listed in table~\ref{tab:obs}, both without
applying any cuts, and with the selection cuts of eq.~(\ref{eq:cuts});
all moments are evaluated on the fly (i.e.~not a-posteriori using the
corresponding differential distribution), as explained in 
appendix~\ref{sec:comp}. At the end of the runs, we have the predictions
for the Mellin moments that correspond to the central scales and PDF set,
and to all non-central scales and PDFs that belong to the relevant error set; 
as already explained, all the non-central results do not require additional
runs, but are obtained through reweighting. The envelopes of the non-central
scale and PDF results are then separately constructed. Finally, the scale 
and PDF uncertainties are combined in quadrature. 

\begin{figure}
 \begin{center}
  \epsfig{file=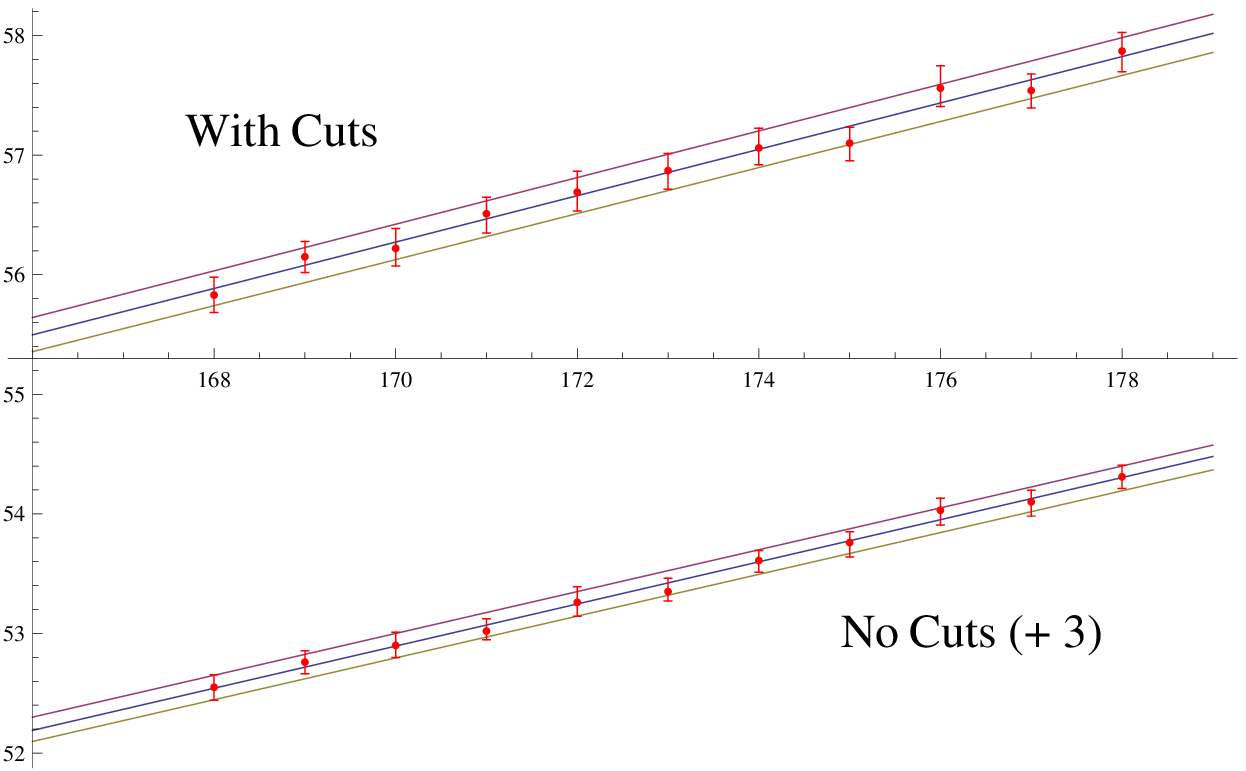,     width=0.45\textwidth}
  \hskip 7mm
  \epsfig{file=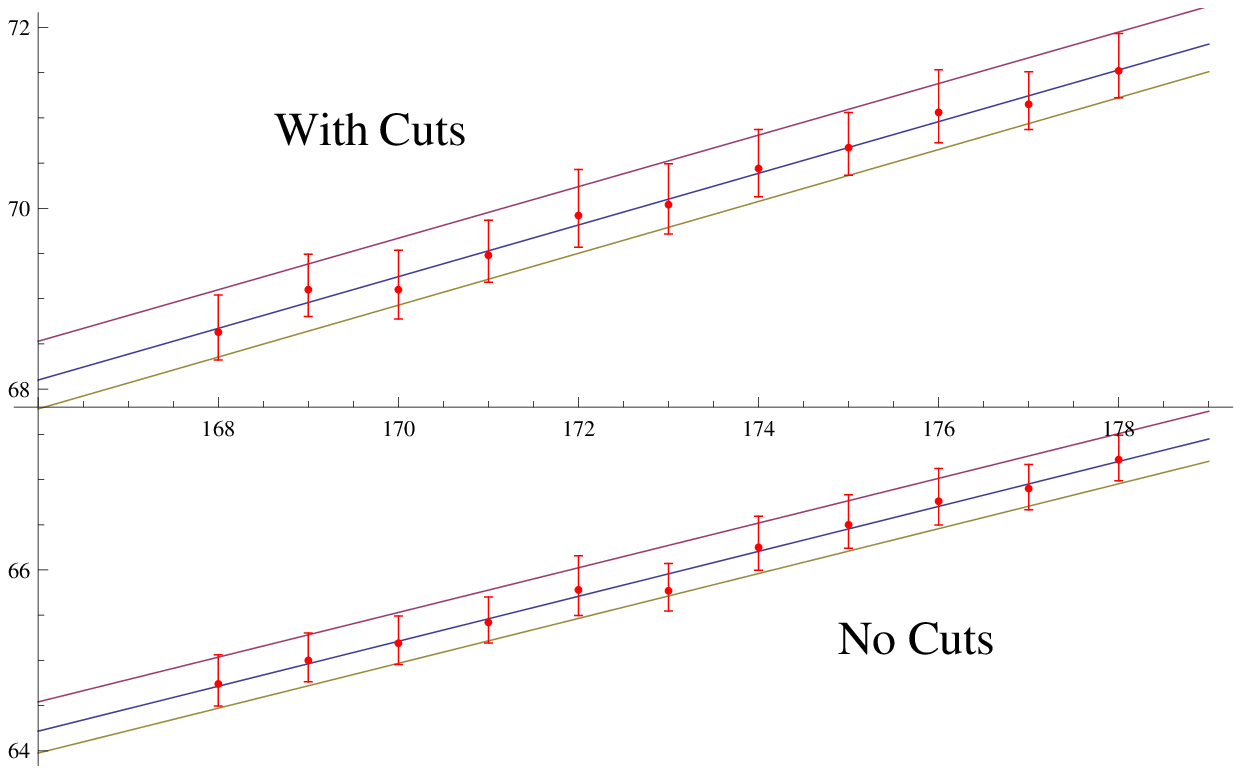, width=0.45\textwidth}
 \end{center}
\caption{Results (red vertical lines) for the first moment of 
$\pt(\ell^+)$ (left panel) and $\pt(\ell^+\ell^-)$ (right panel), 
with and without the cuts of eq.~(\ref{eq:cuts}), in an NLO+PS+MS simulation
with the scale of eq.~(\ref{eq:scalesMtMT}). In each case, the three lines 
are the best straight-line fits to the centres or to the upper/lower ends 
of the theoretical error band, and are thus identified with the functions
$f_{C,U,L}$. The lower (``No Cuts") band on the left panel 
is shifted upwards by $3\GeV$ for improved readibility.}
\label{fig:ptl}
\end{figure}

The bottom line is that at the end of a given run for each Mellin moment
we obtain three numbers: the central, upper, and lower predictions 
for that moment. Examples of such an outcome, for all the $\mtop$
values in the set~(\ref{eq:mtopset}), are given in fig.~\ref{fig:ptl}
in the form of the usual error-bar layout.
Both panels of fig.~\ref{fig:ptl} are relevant to NLO+PS+MS simulations
with the scale choice of eq.~(\ref{eq:scalesMtMT}); the one on the left
(right) reports the first moment of $\pt(\ell^+)$ ($\pt(\ell^+\ell^-)$),
both with and without selection cuts. 

Having the above results, the set of the eleven central, or upper, or lower, 
values for each of the moments is then fitted with the following
functional form:
\beq
\frac{\Mm{O}{i}(\mtop)}{(1~{\rm GeV})^i} = 
\alpha_O^{(i)}\,173^i + 
\beta_O^{(i)}\left(\frac{\mtop}{1~{\rm GeV}}\right)^i\,.
\label{eq:res-equation}
\eeq
The best fits to the central, upper, and lower moments are finally 
identified with the functions $f_C$, $f_U$, and $f_L$, respectively, introduced
in sect.~\ref{sec:extract}. In the examples of fig.~\ref{fig:ptl}, these
three functions are the three straight lines (for each of the four situations
considered there); the analogy with fig.~\ref{fig:extract} is evident.
The actual fitted values of the coefficients of eq.~(\ref{eq:res-equation})
that correspond to the $f_{C,U,L}$ functions in the case of the selection
cuts, together with their analogues for the other three observables not
shown in fig.~\ref{fig:ptl},
are reported in table~\ref{tab:res-coeff}. They will not be explicitly used 
in what follows, and simply constitute a benchmark for future applications.

\begin{table}[h]
\begin{center}
\begin{tabular}{c|c|c|c|c|c|c}
Observable & $\alpha$ (central) & $\beta$ (central) & $\alpha$ (upper) & $\beta$ (upper) & $\alpha$ (lower) & $\beta$ (lower) 
\\\hline
$\pt(\ell^+)$ & 0.1347 & 0.1939 & 0.1345 & 0.1950 & 0.1353 & 0.1925 \\
$\pt(\ell^+\ell^-)$ & 0.1195 & 0.2857 & 0.1227 & 0.2850 & 0.1166 & 0.2868 \\
$M(\ell^+\ell^-)$ & 0.3182 & 0.3109 & 0.3246 & 0.3073 & 0.3216 & 0.3046 \\
$E(\ell^+)+E(\ell^-)$ & 0.5752 & 0.5100 & 0.5663 & 0.5258 & 0.5636 & 0.5154 \\
$\pt(\ell^+)+\pt(\ell^-)$ & 0.2824 & 0.3755 & 0.2896 & 0.3701 & 0.2765 & 0.3796 \\ 
\end{tabular}
\end{center}
\caption{\label{tab:res-coeff}
Values of the coefficients of eq.~(\ref{eq:res-equation}), for the 
first moments resulting from an NLO+PS+MS simulation
with the scale of eq.~(\ref{eq:scalesMtMT}); the selection cuts
of eq.~(\ref{eq:cuts}) are applied.
}
\end{table}
We conclude this section by pointing out that statistical integration
errors are completely neglected in the fitting procedure described
above (which is equivalent to taking all of them equal). In fact,
the main reason behind choosing such a large number (11) of $\mtop$
values for our simulations is that of minimising the impact of statistical
fluctuations, without having to bother about the statistical errors,
which are notoriously tricky in the case of the integration of 
NLO cross sections. The typical size of the statistical fluctuations
can be gathered from fig.~\ref{fig:ptl}; it tends to increase in the
case of higher moments, but it remains manageable up to the fourth
moment, which is the largest we have considered. Obviously, statistical
fluctuations can be reduced by increasing the accuracy of all runs
performed. Given the large number of simulations relevant to the
present paper, we have limited ourselves to work with $10^6$ events
(of which, about $30\%$ pass the selection cuts of eq.~(\ref{eq:cuts}))
in the case of computations matched to partons showers, and with a
comparable accuracy in the case of fixed-order calculations.
With this setup, we have found that over the interval $168-178$ GeV the
functional form of eq.~(\ref{eq:res-equation}) gives an excellent
fit up to the fourth moment, and we believe that this conclusion
applies regardless of the statistics; in other words, we see no
reason for considering polynomials of higher orders in $\mtop^i$ 
in the fitting procedure.

\subsection{Extraction of the top quark mass}
We now use the predictions for $f_{C,U,L}(\mtop)$, calculated as 
described in sect.~\ref{sec:thsetups}, to extract the value
of the top quark mass according to the procedure outlined in
sect.~\ref{sec:extract}. In this way, we shall obtain the main
figure of merit relevant to the method proposed in this paper,
namely the size of the theoretical errors, eq.~(\ref{therr}),
associated with the extraction. In addition, by comparing the 
results emerging from the different computational scenarios of
table~\ref{tab:calc}, we shall assess the presence and numerical
impact of the possible theory biases that affect the various 
approximations.

In order to carry out the programme just described, we need to
start from some data, as in eq.~(\ref{mudata}). In view of the
fact that Mellin moments for the observables of table~\ref{tab:obs}
have not been measured at the LHC, we shall generate them ourselves,
by using the procedure to be described in sect.~\ref{sec:pd}. We point
out that theoretically-generated (pseudo)data are actually more 
advantageous than real data if one is interested in studying the
performances of a given procedure, since they provide one with a
fully-controlled environment.

All of the theory predictions and pseudodata used in this section
have been subject to the selection cuts of eq.~(\ref{eq:cuts}).

\subsubsection{Pseudodata\label{sec:pd}}
Since we believe that our reference scenario, namely NLO+PS+MS,
will give the best description of actual (SM) physics, it is natural
to adopt it for the generation of the pseudodata. While well-motivated
from a physics viewpoint, we stress that, for the sake of a purely
theoretical exercise, this choice is completely arbitrary, and that
the conclusions we shall arrive at would be qualitatively unchanged
had we chosen a different scenario.
The pseudodata are generated by setting:
\beqn
&& \mtoppd=174.32~\GeV\,, 
\nonumber\\
&& \muF^{\rm pd} =  0.45\left(m_{{\sss T},t}+m_{{\sss T},\bar t}+
0.5p_{{\sss T},p}\right)\,,
\nonumber\\
&& \muR^{\rm pd} = 0.60\left(m_{{\sss T},t}+m_{{\sss T},\bar t}+
0.3p_{{\sss T},p}\right)\,,
\label{eq:pd}
\eeqn
where in the definitions of the scales use is made of the transverse 
masses of the top and the antitop, and of the transverse momentum of 
the massless parton possibly present in the final state at the 
hard-subprocess level.

With the choices in eq.~(\ref{eq:pd}), pseudodata generation does not
correspond to any of the scenarios of table~\ref{tab:calc} and to any 
of the scales of eqs.~(\ref{eq:scalesMtMT})--(\ref{eq:scalesmt}).
Therefore, one must not expect that the extractions of the top mass
will return exactly $\mtoppd$, owing to the presence of the biases
discussed in sect.~\ref{sec:bias}. Having said that, we expect pseudodata
to show a clear ``preference'' (i.e.~smaller biases) towards simulations
based on the NLO+PS+MS scenario, since in those cases the biases must be 
due only to scale choices. The verification that this is indeed the case will
constitute a self-consistency check of the procedure we are following.
Note that the information relevant to theory biases is encoded 
not in the actual value of the extracted $\mtop$, but in its difference
with $\mtoppd$. Because of the behaviour of the $f_{C,U,L}$ functions,
such a difference is very much insensitive to the choice of $\mtoppd$,
which allows one to pick an arbitrary value for the latter, as is
done in eq.~(\ref{eq:pd}), and which is ultimately the reason for
the robustness of the usage of pseudodata.

\subsubsection{Shower, NLO, and spin-correlation effects\label{sec:eff}}
The scenarios of table~\ref{tab:calc} differ by the various approximations
they are based upon, each of which may lead to biases in the extraction of
the top mass. An interesting question is then whether the different
sources of possible biases can be disentangled from each other 
(i.e.~whether in a sense they factorise). This is not only relevant
in the context of the present exercise, but also because it may help
assess the impact of approximations not considered here (such as
NNLO corrections), and which might become crucial in the presence
of real data. Furthermore, it also sheds light on the characteristics
of the various observables used in this paper, and in so doing suggests
how to enlarge their set.

In order to address the items above, we proceed as follows.
We select pairs of scenarios that differ in one and only one aspect
of the approximations they involve; for example, scenarios \#1 and \#2
differ in the perturbative accuracy (NLO vs LO) of the underlying
computations. The aspects that we shall be able to consider are
three, namely parton-shower, NLO-correction, and spin-correlation effects,
which we shall discuss in turn below. The top mass extracted
within scenario \#$i$ will be denoted by:
\beq
\mtop^{(i)}\,.
\eeq
Let us then suppose to have chosen a pair of scenarios $(\#i,\#j)$
that differ only by aspect {\bf A}. What we may consider are the
quantities:
\beqn
&&\mtop^{(i)}-\mtoppd\,,\;\;\;\;\;\;\;\;
\mtop^{(j)}-\mtoppd\,,
\label{ijmpd}
\\
&&\mtop^{(i)}-\mtop^{(j)}\,.
\label{imj}
\eeqn
While the differences in eq.~(\ref{ijmpd}) are sensitive to all theory 
biases that affect scenarios \#$i$ and \#$j$, we expect that the
difference in eq.~(\ref{imj}) is solely sensitive to the effect of {\bf A}
(if the factorisation property mentioned above holds to some extent).
In the following, we report the differences that appear
in eq.~(\ref{ijmpd}) and~(\ref{imj})\footnote{Owing to the linear dependence
of these three quantities, only one of the differences in eq.~(\ref{ijmpd}) 
will be shown.}, for all the relevant $(\#i,\#j)$ pairs and all the observables
of table~\ref{tab:obs}. We shall limit ourselves to considering the 
first moments, which are sufficient for the sake of the present exercise; 
all results are obtained with the scale of eq.~(\ref{eq:scalesMtMT}).
In the case of eq.~(\ref{imj}), which is our main interest
here, we also report the errors affecting the difference, which is 
computed by combining in quadrature the errors (determined according
to eq.~(\ref{therr})) that affect the individual $\mtop^{(i)}$ and
$\mtop^{(j)}$ values. The errors on the differences in eq.~(\ref{ijmpd})
are of comparable size, up to a factor $\sqrt{2}$ smaller since
$\mtoppd$ is assumed to be known with infinite precision.

We start with shower effects, and report the corresponding results
in table~\ref{tab:pulls-shower}. The relevant scenario pairs are
$(3,5)$ and $(4,6)$, the latter being the LO counterpart of the
former, which is accurate to NLO. Note that scenarios \#1 and \#2
have not been considered here, owing to the lack of fixed-order
results that include production spin correlations.
\begin{table}[h]
\begin{center}
\begin{tabular}{c|c|c||c|c}
obs. & $\mtop^{(3)} - \mtop^{(5)}$ & $\mtop^{(3)} - \mtoppd$ & $\mtop^{(4)} - \mtop^{(6)}$ & $\mtop^{(4)} - \mtoppd$
\\\hline
1 & $-0.35^{+1.14}_{-1.16}$ &  $+0.12$ 		&    $-2.17^{+1.50}_{-1.80}$ & $-0.67$ \\
\hline
2 & $-4.74^{+1.98}_{-3.10}$ & $+11.14$ 		&    $-9.09^{+0.76}_{-0.71}$ & $+14.19$ \\
\hline
3 & $+1.52^{+2.03}_{-1.80}$ & $-8.61$ 		&     $+3.79^{+3.30}_{-4.02}$ & $-6.43$ \\
\hline
4 & $+0.15^{+2.81}_{-2.91}$ & $-0.23$ 		&    $-1.79^{+3.08}_{-3.75}$ & $-1.47$ \\
\hline
5 & $-0.30^{+1.09}_{-1.21}$ & $ +0.03$ 		&    $-2.13^{+1.51}_{-1.81}$ & $-0.67$ \\
\end{tabular}
\end{center}
\caption{\label{tab:pulls-shower} 
Impact of parton showers on mass extractions. See the text for details.}
\end{table}
The first observation is that the $(3,5)$ and $(4,6)$ cases are rather
consistent with each other; however, the results for eq.~(\ref{imj})
of the latter are in absolute value systematically larger than
those of the former. This is compatible with the expectation,
corroborated by ample heuristic evidence in many different processes,
that shower effects are milder if the underlying computations are
NLO-accurate (as opposed to LO ones), for the simple reason that 
NLO results do already include part of the radiation to be generated
by parton showers\footnote{This also shows that NLO and shower effects do 
not factorise entirely; it remains true that they affect the $\mtop$
extraction for a given observable in different manners.}. While in the case 
of NLO-based simulations all differences are statistically compatible with 
zero (within $1\sigma$) except for observable \#2, in the case of LO-based 
simulations more significant deviations can be seen in the cases of 
observables \#1 and \#5 as well. The take-home message, then,
is that shower effects are moderate if higher-order corrections
are taken into account, which is good news in view of the future
availability of NNLO parton-level differential results; however,
this conclusion does not apply to the transverse momentum of the
charged-lepton pair, for which a proper matching with parton
showers appears to be needed.

As far as the results for eq.~(\ref{ijmpd}) are concerned,
table~\ref{tab:pulls-shower} shows that values significantly
different from zero are obtained in the cases of observables \#2
and \#3. The size of the difference relevant to \#2 is larger
than that resulting from eq.~(\ref{imj}), which implies that
for such an observable other effects, on top of those due to showers,
are sources of theory biases as well (both NLO and spin correlations, 
as we shall show later). A similar conclusion applies to the
lepton-pair invariant mass \#3, for which the absence of shower
effects implies that biases are entirely due to some other 
mechanism (spin correlations, as it will turn out).

\begin{table}[h]
\begin{center}
\begin{tabular}{c|c|c||c|c||c|c}
obs. & $\mtop^{(1)} - \mtop^{(2)}$ & $\mtop^{(1)} - \mtoppd$ & $\mtop^{(3)} - \mtop^{(4)}$ & $\mtop^{(3)} - \mtoppd$ & $\mtop^{(5)} - \mtop^{(6)}$ & $\mtop^{(5)} - \mtoppd$
\\\hline
1 &	$+1.16^{+1.43}_{-1.60}$ & $+0.41$		&	$+0.79^{+1.43}_{-1.60}$ & $+0.12$		&	$-1.03^{+1.22}_{-1.43}$ & $+0.47$ \\\hline
2 &   $-2.79^{+1.27}_{-1.65}$ & $-1.18$		&	$-3.05^{+1.35}_{-1.64}$ & $+11.14$		&	$-7.41^{+1.64}_{-2.72}$ & $+15.87$ \\\hline
3 &   $-0.73^{+3.21}_{-3.45}$ & $+0.84$		&	$-2.18^{+3.03}_{-3.30}$ & $-8.61$		&	$+0.09^{+2.42}_{-2.91}$ & $-10.13$ \\\hline
4 &    $+1.74^{+3.27}_{-3.78}$ & $+0.16$		&	$+1.23^{+3.10}_{-3.61}$ & $-0.23$		&	$-0.70^{+2.79}_{-3.09}$ & $-0.38$ \\\hline
5 &    $+0.99^{+1.42}_{-1.72}$ & $+0.25$		&	$+0.70^{+1.40}_{-1.72}$ & $+0.03$		&	$-1.13^{+1.23}_{-1.33}$ & $+0.33$ \\
\end{tabular}
\end{center}
\caption{\label{tab:pulls-nlo} 
Impact of NLO corrections on mass extractions. See the text for details.}
\end{table}
We next consider NLO effects, which we document in table~\ref{tab:pulls-nlo},
and for which the relevant scenario pairs are $(1,2)$, $(3,4)$, and $(5,6)$.
As far as eq.~(\ref{imj}) is concerned, the differences for all pairs
and all observables except \#2 are compatible with zero; thus, the
first moments of such observables appear to be quite stable perturbatively,
regardless of the matching to parton showers, and of the presence of
spin correlations. For what concerns $\pt(\ell^+\ell^-)$, on top of
the fact that NLO effects are significant in all scenarios, we observe
that they are particularly strong when the matching to showers is
not performed (pair $(5,6)$); this is again related to the fact that,
in certain corners of the phase space, showers and NLO corrections
affect the kinematics in a similar way. Coming to the absolute size
of theory biases, eq.~(\ref{ijmpd}), we see that they are all rather
small in the case of NLO+PS+MS predictions (second column); this is
what we expect, as explained in sect.~\ref{sec:pd}. For the other
scenarios, large differences are observed in the case of observables
\#2 and \#3, which was expected in view of table~\ref{tab:pulls-shower}.
For the latter observable, this fact, the absence of NLO effects,
and the results of table~\ref{tab:pulls-shower} imply that the biases
are solely due to spin correlations.

\begin{table}[h]
\begin{center}
\begin{tabular}{c|c|c||c|c}
obs. & $\mtop^{(1)} - \mtop^{(3)}$ & $\mtop^{(1)} - \mtoppd$ & $\mtop^{(2)} - \mtop^{(4)}$ & $\mtop^{(2)} - \mtoppd$
\\\hline
1 & $+0.29^{+1.17}_{-1.14}$ & $+0.41$		&	$-0.08^{+1.66}_{-1.96}$ & $-0.75$ \\\hline
2 & $-12.32^{+1.62}_{-2.13}$ & $-1.18$		&	$-12.58^{+0.90}_{-0.94}$ & $+1.60$ \\\hline
3 & $+9.45^{+2.36}_{-2.16}$ & $+0.84$		&	$+8.00^{+3.74}_{-4.26}$ & $+1.57$ \\\hline
4 & $+0.39^{+2.93}_{-3.16}$ & $+0.16$		&	$-0.11^{+3.42}_{-4.16}$ & $-1.58$ \\\hline
5 & $+0.22^{+1.12}_{-1.28}$ & $+0.25$		&	$-0.06^{+1.65}_{-2.07}$ & $-0.73$ \\
\end{tabular}
\end{center}
\caption{\label{tab:pulls-spincorr} 
Impact of spin correlations on mass extractions. See the text for details.}
\end{table}
We finally turn to spin-correlations effects, whose results are
reported in table~\ref{tab:pulls-spincorr}, and for which the 
relevant scenario pairs are $(1,3)$ and $(2,4)$; these two pairs
differ in the underlying perturbative accuracy, which is NLO and LO
respectively. The conclusions that can be drawn from 
table~\ref{tab:pulls-spincorr} have already been anticipated.
Namely, that the differences resulting from eq.~(\ref{imj}) are
significantly different from zero for both observables \#2 and \#3,
while they are negligible in the other cases. The sizes of the former
differences appear to be fairly insensitive to NLO corrections,
which is an indirect confirmation of the factorisation of spin-correlation
effects.

The general conclusion of this section is the following. Observables
that are single-inclusive ($\pt(\ell^+)$), and that feature a mild
correlation between the decay products of the top and antitop
($E(\ell^+)+E(\ell^-)$ and $\pt(\ell^+)+\pt(\ell^-)$), are rather
stable against shower, NLO, and spin-correlations effects. This
is not true for observables for which the correlation between the
two charged leptons is stronger ($\pt(\ell^+\ell^-)$ and $M(\ell^+\ell^-)$):
the fact that either shower or spin-correlation effects (or both)
are relevant implies, among other things, that the computation of the 
$t\bt$ cross section at the NNLO with stable tops will not be sufficient to 
give a good description of such observables, at the very least in 
the context of the top mass extraction considered in this paper.

\subsubsection{Results for the top quark mass\label{sec:res}}
In this section we present the results for the extraction of
the top quark mass obtained with our reference computational
scenario, NLO+PS+MS. We are specifically interested in
checking the size of the theory uncertainty affecting such
an extraction, and its behaviour (together with that of the central
top quark mass) when the results emerging from the individual
observables and moments are combined together. These findings
will also serve as benchmarks for the studies that we shall carry
out in sect.~\ref{sec:resbias}, where the extraction of the 
top mass will be performed by using the other scenarios
of table~\ref{tab:calc}. Furthermore, we want to study how the 
above results are influenced by the scale choice, and therefore we 
shall consider all of the three forms given in 
eqs.~(\ref{eq:scalesMtMT})--(\ref{eq:scalesmt}).

The general strategy is the following. For a given scale choice, we 
extract the top mass from each of the five observables of table~\ref{tab:obs}
and their first three moments\footnote{The fourth moments turn out
not to be particularly useful in the extraction procedure, being
affected by errors larger than those of the lower moments, and being
rather strongly correlated with the third moments; these are the reasons 
why they are not taken into account.}, i.e.~fifteen $\mtop$ values 
in total, each with its theory errors of eq.~(\ref{therr}).
These values, or any subset of them, are then combined to obtain
the ``best'' result. The combination technique is briefly explained
in appendix~\ref{sec:comb}, and is rather standard: basically, the
central values are weighted with the inverse of the square of their
errors. Since the various observables and their moments are correlated,
it is necessary to take these correlations into account, lest one
skew the final central value of $\mtop$ and underestimate its error.

The simplest case is that where one uses a single observable for
extracting $\mtop$; as was explained in sect.~\ref{sec:bias},
this is far from being ideal, and we present it here only as
a way to compare with the multi-observable results that will
be shown later. We use observable \#1 ($\pt(\ell^+)$) because
it is the one whose top-mass extractions are affected by the
smallest errors (in the case of the scale of 
eq.~(\ref{eq:scalesMtMT})). The values of $\mtop$ that we obtain are
given in table~\ref{tab:nlopsms-1}, which should be read as follows
(this layout will be used for the other tables of this section
as well). Each one of the first three rows corresponds to one of the
scales of eqs.~(\ref{eq:scalesMtMT})--(\ref{eq:scalesmt})
(i.e.~the $i^{th}$ row is obtained with $\hat\mu^{(i)}$).
The first, second, and third column reports the results obtained
by considering only the first, up to the second, and up to the third
Mellin moments, respectively. The results in the fourth row are obtained
by combining the three results that appear in the first three rows
of the same column. Such a combination is achieved by weighting those
three results with the inverse of the square of their errors. Since the
errors are asymmetric, one treats separately the $+$ and $-$ ones;
the two resulting ``central'' $\mtop$ values are possibly different,
and the single $\mtop$ reported in table~\ref{tab:nlopsms-1} is then
obtained again with a weighted average. Finally, the numbers in
square brackets are the values of $\chi^2$ per degree of freedom,
computed by always considering the first four Mellin moments, 
regardless of how many of them had been actually used in the 
combination. One should not seek a deep meaning in
this $\chi^2$, in particular because of the way the errors that
enter into it are obtained (i.e.~their behaviour from a statistical
viewpoint is unknown to us). On the other hand, while its precise value
is not of particular significance, it represents a very useful reference for 
the performance of the extraction procedure, as we shall see in
sect.~\ref{sec:resbias}.
\begin{table}[h]
\begin{center}
\begin{tabular}{cccc}
scale & $i=1$ & $i=1\plus 2$ & $i=1\plus 2\plus 3$
\\\hline
1 & $174.73^{+0.80}_{-0.79}[0.2]$ & $174.73^{+0.80}_{-0.79}[0.2]$ & $174.72^{+0.80}_{-0.79}[0.2]$ \\
2 & $174.78^{+0.90}_{-0.90}[0.6]$ & $174.78^{+0.90}_{-0.90}[0.6]$ & $174.78^{+0.90}_{-0.90}[0.6]$ \\
3 & $172.73^{+2.0}_{-1.2}[0.5]$ & $172.73^{+1.96}_{-1.19}[0.5]$ & $172.73^{+1.96}_{-1.19}[0.5]$ \\
$1\plus 2\plus 3$ & $174.46^{+0.99}_{-0.92}$ & $174.46^{+0.99}_{-0.92}$ & $174.45^{+0.99}_{-0.92}$ \\
\end{tabular}
\end{center}
\caption{\label{tab:nlopsms-1} 
Top mass values extracted from observable \#1, with up to three moments,
and for three different scale choices. The last line reports the results
obtained by combining the central $\mtop$ values relevant to the three 
scales. The numbers in square brackets are $\chi^2/n$. The pseudodata
top mass is $\mtoppd=174.32$~GeV. See the text for details.}
\end{table}

The messages to be taken out of table~\ref{tab:nlopsms-1} are the following.
Firstly, the impact of the addition of moments beyond the first is extremely
modest, if visible at all. This is due to the fact that the errors
affecting $\mtop$ increase with higher moments, and to the non-negligible
correlations between the moments (see appendix~\ref{sec:comb}).
Secondly, the scales $\hat\mu^{(1)}$ and $\hat\mu^{(2)}$ tend to give central
results larger than the ``true'' one of the pseudodata, $\mtoppd=174.32~\GeV$,
while the opposite applies to scale $\hat\mu^{(3)}$, where the 
effect is more evident (but still within $1\sigma$). Let us then consider
the latter case to be definite, and compare the functional form of 
eq.~(\ref{eq:scalesmt}) with those of eq.~(\ref{eq:pd}). Because of
the dependence on the transverse momenta of the scales used in the
pseudodata, which is absent in the case of $\hat\mu^{(3)}$, the tails
of the $\pt$-related distributions obtained with $\hat\mu^{(3)}$
will be less rapidly falling than those of the pseudodata (mainly 
because the $\pt$-dependence of $\muR$ in eq.~(\ref{eq:pd}) will
induce a stronger $\as$ suppression, relative to the small-$\pt$ region,
than in the case of $\hat\mu^{(3)}$; this effect is only mildly compensated
by that due to $\muF$). Thus, the moments computed with scale \#3
will be slightly larger than their analogues in the pseudodata.
For the reasons explained in sect.~\ref{sec:bias}, this difference then
results in a lower (than the input $\mtoppd$) value for the extracted
top mass, which is what we see in the third row of table~\ref{tab:nlopsms-1}.
The same effect, but (slightly) in the opposite direction, is at play in the 
case of scales \#1 and \#2. Here, the numerical values of such scales at
large $\pt$'s relative to their small $\pt$ counterparts
are closer to those relevant to the pseudodata scales than in
the case of scale \#3, whence closer-to-$\mtoppd$ central results
for the top mass. Given these opposite behaviours, not surprisingly
the average of the three results is closer to $\mtoppd$ than any of
them; such an average is biased towards the results of $\hat\mu^{(1)}$
and $\hat\mu^{(2)}$, owing to their errors being smaller than those
associated with the extractions with $\hat\mu^{(3)}$.

\begin{table}[h]
\begin{center}
\begin{tabular}{ccccc}
scale & $i=1$ & $i=1\plus 2$ & $i=1\plus 2\plus 3$
\\\hline
1 & $174.67^{+0.75}_{-0.77}[3.0]$ & $174.67^{+0.75}_{-0.77}[3.0]$ & $174.61^{+0.74}_{-0.77}[3.2]$ \\
2 & $174.81^{+0.83}_{-0.80}[6.2]$ & $174.80^{+0.82}_{-0.80}[6.2]$ & $174.85^{+0.82}_{-0.80}[6.1]$ \\
3 & $172.63^{+1.85}_{-1.16}[0.2]$ & $172.64^{+1.82}_{-1.15}[0.2]$ & $172.58^{+1.81}_{-1.15}[0.2]$ \\
$1\plus 2\plus 3$ & $174.44^{+0.92}_{-0.87}$ & $174.44^{+0.92}_{-0.87}$ & $174.43^{+0.91}_{-0.87}$ \\
\end{tabular}
\end{center}
\caption{\label{tab:nlopsms-145} 
As in table~\ref{tab:nlopsms-1}, with the extractions performed by
using observables \#1, \#4, and \#5. The pseudodata top mass is 
$\mtoppd=174.32$~GeV.}
\end{table}
We now repeat the combination procedure that has led to the results of 
table~\ref{tab:nlopsms-1}, by including, on top of the $\mtop$ values 
obtained with observable \#1, also those relevant to observables \#4 and \#5; 
the new combined results are presented in table~\ref{tab:nlopsms-145}.
By far and large, all comments relevant to table~\ref{tab:nlopsms-1}
can be repeated here. There is a decrease (less than 10\% for all scales) 
of the errors, which is not large because of two facts: observable \#1
induces the smallest errors (in the present observable set), 
and the observables considered are 
sizably correlated, as documented in appendix~\ref{sec:comb}.
By adding more observables one starts to see the effects of the
inclusion of higher moments; although statistically not significant,
there are trends in the central values which were not visible in
the case of a single observable.

\begin{table}[h]
\begin{center}
\begin{tabular}{ccccc}
scale & $i=1$ & $i=1\plus 2$ & $i=1\plus 2\plus 3$
\\\hline
1 & $174.48^{+0.73}_{-0.77}[5.0]$ & $174.55^{+0.72}_{-0.76}[5.0]$ & $174.56^{+0.71}_{-0.76}[5.1]$ \\
2 & $174.73^{+0.77}_{-0.80}[4.3]$ & $174.74^{+0.76}_{-0.79}[4.3]$ & $174.91^{+0.75}_{-0.79}[4.1]$ \\
3 & $172.54^{+1.03}_{-1.07}[1.6]$ & $172.46^{+0.99}_{-1.05}[1.6]$ & $172.22^{+0.95}_{-1.04}[1.4]$ \\
$1\plus 2\plus 3$ & $174.16^{+0.81}_{-0.85}$ & $174.17^{+0.80}_{-0.84}$ & $174.17^{+0.78}_{-0.84}$ \\
\end{tabular}
\end{center}
\caption{\label{tab:nlopsms-all} 
As in table~\ref{tab:nlopsms-1}, with the extractions performed by
using all observables. The pseudodata top mass is $\mtoppd=174.32$~GeV.}
\end{table}
Finally, in table~\ref{tab:nlopsms-all} we present the results obtained
by combining the extractions of $\mtop$ from all observables; thus,
according to the discussion given in sect.~\ref{sec:bias}, these have 
to be considered our best estimates of the top mass, given the
pseudodata of sect.~\ref{sec:pd}. The errors decrease further 
w.r.t.~those of table~\ref{tab:nlopsms-145} (not significantly in 
the case of $\hat\mu^{(1)}$ and $\hat\mu^{(2)}$, but by a large factor 
for $\hat\mu^{(3)}$; this is because, for such a scale, it is the
$\pt$ of the lepton pair that happens to be affected by the smallest
errors). The trend induced by the addition of higher moments becomes 
more visible than before, and statistically significant (a
$2\sigma$ effect) in the case of scale \#3. However, the final
results of the fourth row, obtained by combining the outcomes
associated with the different scales, are quite stable. The case
of the results associated with $\hat\mu^{(3)}$ is interesting, because
it stresses again the importance of considering as many observables
and as diverse as possible in order to expose potential theory biases
in the top-mass extraction. Given that here all our predictions are based
on the same computational scenario as the pseudodata, namely NLO+PS+MS,
the only deviations from a perfect reconstruction can only be due
to the different choice of scales, and $\hat\mu^{(3)}$ happens to be
farther from the pseudodata ones of eq.~(\ref{eq:pd}) than either
$\hat\mu^{(1)}$ or $\hat\mu^{(2)}$. The crucial point is that this
observation is true regardless of the type of observables considered,
but it is only when the lepton-pair correlations \#2 and \#3 enter 
the combination that the effects become more noticeable. This is related
to the behaviour of these two observables discussed in sect.~\ref{sec:eff},
which exhibit the strongest sensitivity to (among other things) extra
radiation. A change of scale is an effective, if quite mild, way of
probing some of these extra-radiation effects. As we shall see in
sect.~\ref{sec:resbias}, the impact of the addition of these two
observables on the theory biases is spectacular when the underlying
calculational scenario is different w.r.t.~that used in the generation
of the pseudodata.

There are two conclusions that can be drawn from this section.
The first is that the procedure proposed in this paper appears
to be able to give theory errors on the extracted top mass of the 
order of $0.8$~GeV. While we have neglected background contaminations,
we have also been conservative with the range of scale variations;
on top of that, the addition of further observables may help reduce
further those errors. The second conclusion is more general, in that
it applies to any extraction method based on templates. Our exercise 
demonstrates that one thing is the variation of the scales induced by 
pre-factors that multiply a given functional form, and quite another
the change of that functional form. Although the two procedures overlap,
they are not equivalent. We have shown a practical way to probe the 
changes of the above functional form: the idea is that, by re-computing
theoretical predictions for many different scale choices, and by
performing a weighted average of their outcomes, one might effectively
capture the scale settings which optimally describe Nature.

\subsubsection{More on theory biases\label{sec:resbias}}
The aim of this section is that of repeating what has been
done in sect.~\ref{sec:res}, for scenarios other than NLO+PS+MS.
In other words, all of the computations considered here are 
different w.r.t.~that used in the generation of the pseudodata;
we shall thus study the theory biases, whose sources we have already 
discussed in sect.~\ref{sec:eff}, at the level of the combined results 
for the extracted top quark mass. All the calculations are performed
by using the scale $\hat\mu^{(1)}$. We report the results in 
table~\ref{tab:results-rest}, which is organised with the same
conventions as those used in the tables of sect.~\ref{sec:res}.
This table is split into two parts, relevant to the $\mtop$ 
extraction performed by using only three observables (\#1, \#4, 
and \#5), or all of them. These two parts thus are in one-to-one
correspondence with (the first row of) tables~\ref{tab:nlopsms-145} 
and~\ref{tab:nlopsms-all}, respectively.

\begin{table}[h]
\begin{center}
\begin{tabular}{cccc}
Scenario & $i=1$ & $i=1\plus 2$ & $i=1\plus 2\plus 3$
\\\hline\hline
 & \multicolumn{3}{c}{Observables \#1, \#4, \#5} \\
\hline
 LO+PS+MS & $173.61^{+1.10}_{-1.34}[1.0]$ & $173.63^{+1.10}_{-1.34}[1.0]$ & $173.62^{+1.10}_{-1.34}[1.0]$ \\
 NLO+PS & $174.40^{+0.75}_{-0.81}[3.5]$ & $174.43^{+0.75}_{-0.81}[3.5]$ & $174.60^{+0.75}_{-0.79}[3.2]$ \\
 LO+PS & $173.68^{+1.08}_{-1.31}[0.8]$ & $173.68^{+1.08}_{-1.31}[0.9]$ & $173.75^{+1.08}_{-1.31}[0.9]$ \\
 fNLO & $174.73^{+0.72}_{-0.74}[5.5]$ & $174.72^{+0.71}_{-0.74}[5.6]$ & $175.18^{+0.64}_{-0.71}[4.6]$ \\
 fLO & $175.84^{+0.90}_{-1.05}[1.2]$ & $175.75^{+0.89}_{-1.05}[1.2]$ & $175.82^{+0.89}_{-1.04}[1.2]$ \\
\hline\hline
 & \multicolumn{3}{c}{All observables} \\
\hline
 LO+PS+MS & $175.98^{+0.63}_{-0.69}[16.9]$ & $176.05^{+0.63}_{-0.68}[17.8]$ & $176.12^{+0.61}_{-0.68}[18.9]$ \\
 NLO+PS & $175.43^{+0.74}_{-0.80}[29.2]$ & $176.20^{+0.73}_{-0.79}[30.1]$ & $175.67^{+0.73}_{-0.76}[31.2]$ \\
 LO+PS & $187.90^{+0.6}_{-0.6}[428.3]$ & $187.71^{+0.60}_{-0.60}[424.2]$ & $187.83^{+0.58}_{-0.60}[442.8]$ \\
 fNLO & $174.41^{+0.72}_{-0.73}[96.6]$ & $174.82^{+0.71}_{-0.73}[93.1]$ & $175.44^{+0.70}_{-0.68}[94.8]$ \\
 fLO & $197.31^{+0.42}_{-0.35}[2496.1]$ & $197.19^{+0.42}_{-0.35}[2505.6]$ & $197.48^{+0.36}_{-0.35}[3005.6]$ \\
\end{tabular}
\end{center}
\caption{\label{tab:results-rest} 
Combined extracted values of $\mtop$, for various scenarios and two choices
of the set of observables. The pseudodata top mass is $\mtoppd=174.32$~GeV.}
\end{table}
From the upper part of table~\ref{tab:results-rest}, we see that 
the use of observables \#1, \#4, and \#5 leads to central $\mtop$
values which may not be in perfect agreement with the pseudodata
value $\mtoppd$, but are not far from it either, irrespective of
the calculational scenario considered. Furthermore, both the errors
and the $\chi^2$ values are totally reasonable, and rather consistent
with those of table~\ref{tab:nlopsms-145}. These findings need not
be surprising, because they could be anticipated in sect.~\ref{sec:eff},
where observables \#1, \#4, and \#5 have been shown to
be fairly insensitive to shower, NLO, and spin-correlation effects.
These effects are ultimately the difference between each of the 
scenarios considered here, and our reference one, NLO+PS+MS. 
It is therefore instructive to see what 
happens when observables \#2 and \#3 are used in the extractions
as well (lower part of table~\ref{tab:results-rest}). Not only the differences
among the central results for the extracted top mass are much larger than 
before (and particularly so at the LO in absence of proper spin correlations),
but it is especially the $\chi^2$ values that increase dramatically,
in spite of (and, in a sense, thanks to) the fact that the errors 
remain quite moderate. This is exactly the situation that has been
described in sect.~\ref{sec:bias}: the extraction of $\mtop$ from
individual observables is always acceptable and affected by small
errors; however, if the underlying theoretical description is
incompatible with that of the (pseudo)data, the different results
will be mutually incompatible. A (certainly non-unique) way of
making explicit the presence of such incompatibilities is through the 
computation of a $\chi^2$. The lower part of table~\ref{tab:results-rest}
is thus another, very explicit way of showing why considering a large
number of observables with different characteristics is always beneficial,
in this or in other template-based methods.

A final comment on table~\ref{tab:results-rest}. The errors that
affect the extracted top mass do not follow the usual LO$\to$NLO
reduction pattern, and they need not to. Indeed, the relationship
between the above errors, and those which are usually considered
at the level of rates, is rather indirect. Furthermore, in the 
combination of the results obtained from different observables,
a single $\mtop$ value affected by errors much smaller than the
others will have a very large weight, with the picture being
further complicated by the presence of strong correlations among
the observables studied here. While the particular combination
technique used in this paper (see appendix~\ref{sec:comb}) can certainly
be refined, possibly leading to changes in the central values of 
$\mtop$ and their associated errors, the conclusions reached before
will not change, being based on a few well-understood physics phenomena.

\section{Conclusions\label{sec:concl}}
In this paper we have proposed a procedure for the determination
of the top quark pole mass from dilepton $t\bt$ events. Our main 
proposals and findings are the following:
\begin{itemize}
\item 
We use leptonic single-inclusive and correlation observables, which are 
clean and largely insensitive to the modelling of long-distance effects.
Our method, based on Mellin moments, relies neither on the definition 
of the top quark as a pseudo-particle, nor on its reconstruction.
\item
The quality of the results for $\mtop$ and their reliability improves
by increasing the number of observables and of their moments. It is
important that the observables employed have different sensitivities
to the various mechanisms relevant to $t\bt$ production and decay,
such as higher-order corrections, and shower and spin-correlation effects.
Several theoretical simulations must be used that differ in the choice
of the functional form for the hard scales, and the extracted $\mtop$
values must be combined. Thus, we consider the entry in the rightmost
column and last row of table~\ref{tab:nlopsms-all} as our ``best'' result.
\item
The errors associated with $\mtop$ may underestimate the difference 
between the extracted value and the actual pole mass, in the case of
an inadequate theoretical description of the underlying production
mechanism. A $\chi^2$-type test is effective in identifying the presence
of such biases, {\em provided} that a sufficiently large number of
observables has been employed in the extraction procedure, as is
documented in table~\ref{tab:results-rest}.
\end{itemize}
We stress that the second and third items above apply to any template
method that exploits the shapes of observables for the extraction
of the top quark mass.

The most precise $\mtop$ determination that we have achieved with our
method in the context of the purely-theoretical exercise performed
here is affected by errors of the order of 0.8~GeV. It is probably possible 
to reduce this figure further, by using a set of observables larger than
the one considered in this paper. On the other hand, we have not addressed 
two important aspects which will need to be taken into account in an
extraction of $\mtop$ from real data, namely the contamination due 
to backgrounds, and the systematics due to the choice of the parton 
shower Monte Carlo. For what concerns this Monte-Carlo systematics,
it is worth pointing out that within our approach two different Monte 
Carlos must lead to two separate top quark mass values, which should 
eventually be combined on the basis of their respective errors and
of the results of some $\chi^2$ tests. An interesting, if not particularly
desirable, case is that where two Monte Carlos would lead to 
statistically-incompatible $\mtop$ results, with two small $\chi^2$ 
values similar to each other. This implies that the observables chosen
in the extraction procedure do not constrain well enough the theoretical 
models adopted by the Monte Carlos, and it is thus doubtful which (if any)
of the two $\mtop$ results best describes the ``physical'' pole mass.

While we believe that our approach has many competitive features,
it remains true that the determination of the top quark mass will
benefit from the use of many different techniques. For example,
any BSM physics able to modify in a significant manner the kinematic
distributions w.r.t.~those predicted by the SM may induce large biases 
in template-based $\mtop$ extractions, unless the simulation of such
BSM contribution is also taken into account. In this case, an approach
insensitive to the production dynamics (which thus belongs to the
first class introducted in sect.~\ref{sec:intro}) would offer a
valuable addition; one may mention here the CMS end-point 
method~\cite{Chatrchyan:2013boa}, or the promising energy-peak 
method suggested in ref.~\cite{Agashe:2013eba}, provided that
it could be extended to include NLO QCD corrections to top decays.

The approach we have pursued here has many variants which do not change
its essence. For example, one may start looking into $b$-jet variables
in order to increase the sensitivity to $\mtop$; this has the downside
of introducing a larger dependence on long-distance modelling, and the
balance between these two competing aspects must be carefully addressed.
Conversely, one can try and select dilepton events of opposite flavour
without imposing cuts on the $b$ jets, in order to further reduce
the impact of hadronisation; the problem then becomes that of the
control of the backgrounds. Our method is also immediately applicable
in the context of NNLO simulations. However, for this to be effective,
a proper description of top decays, and in particular one that
incorporates production-spin-correlation effects, must be included.
The matching to parton shower would also be highly desirable.

\begin{acknowledgments}
We are indebted to Rikkert Frederix for his technical support and 
collaboration at an early stage of this work. We thank Fabio Maltoni and
Michelangelo Mangano for several helpful discussions. This work has been 
supported in
part by and performed in the framework of the ERC grant 291377 ``LHCtheory: 
Theoretical predictions and analyses of LHC physics: advancing the precision
frontier". The work of A.~M. is also supported by the UK Science and Technology 
Facilities Council (grants ST/L002760/1 and ST/K004883/1). 
\end{acknowledgments}

\appendix

\section{Computation of  moments in the context of event 
generation\label{sec:comp}}

While  the moments of an observable $O$ can be computed by using the 
result for the differential distribution \mbox{$d\sigma/dO$}, there is
actually a more direct way. During the course of an MC simulation,
the fully-differential cross section is expressed through a set 
of $N$ kinematic configurations (``events") and their associated weights:
\beq
d\sigma\;\leftrightarrow\;\Big\{\kin_k,W_k\Big\}_{k=1}^N\,,
\label{sigvsw}
\eeq
with
\beq
\sigma=\sum_{k=1}^N W_k\,.
\label{norm}
\eeq
Note that the $W_k$'s need not necessarily be equal to each other
(in absolute value); in other words, what follows is valid in the
context of both unweighted and weighted event generation, these
being typically relevant to calculations matched to parton-shower
Monte Carlos and at fixed order, respectively. When one computes
a differential cross section, one evaluates event-by-event the 
value of the observable of interest in the generated kinematic
configuration, $O(\kin_k)$; such a value determines, in turn, the bin 
of the corresponding histogram where the weight $W_k$ must be stored.
In a completely analogous manner, the calculation of the (unnormalised) 
 moments can also be performed on the fly. In order to do so,
for a given observable $O$ one will book a histogram with bins
of width one centered at non-negative integers. When the $k^{th}$
event is generated, one stores the weight:
\beq
W_k\,\times\,\Big(O(\kin_k)\Big)^{\,i}
\label{mufill}
\eeq
in the $i^{th}$ bin of the histogram; this must be done for all bins. By 
using eqs.~(\ref{mmdef}), (\ref{norm}), and~(\ref{mufill}) , one sees that 
at the end of the run the $i^{th}$ bin will be equal to the normalised 
$i^{th}$  moment, times $\sigma$, so that the normalised  moments
themselves can be obtained by dividing the content of each bin by
that of the bin centered at zero. 

We point out that this direct way of computing  moments is exact in the 
$N\to\infty$ limit. On the other hand, the (indirect) calculation which 
uses the result of \mbox{$d\sigma/dO$}
is {\em not} exact even in the $N\to\infty$ limit, unless the limit
of vanishing bin size (in \mbox{$d\sigma/dO$}) 
is taken as well, which is impossible
in the context of an actual simulation, where one thus might have
a residual bin-size inaccuracy. Furthermore, in the case where the range
of the histogram in $O$ does not cover the whole kinematically-accessible
range for such an observable, another inaccuracy affects the indirect
computation. For these reasons, and for its greater simplicity,
in this paper we have always adopted the direct, event-by-event 
method outlined above in the calculation of the  moments. We have checked, 
in the case of the first moments, that the results of the direct computations
are very similar, but not identical, to those obtained a-posteriori 
by using the distributions. It must be stressed that the distributions
we have used cover rather large ranges (up to 400~GeV for observables
\#1, \#2, and \#3, up to 1.2~TeV for observable \#4, and up to 1~TeV 
for observable \#5), and contain 100 bins. Therefore, in the context of 
e.g.~an experimental analysis, where the use of large-size bins is typical 
at large momenta, the risk of inaccuracies affecting the moments computed 
from distributions may be non negligible.

\section{Combination of different top quark mass results\label{sec:comb}}
In this appendix we briefly outline the technique used to combine
different $\mtop$ results and their errors. We denote these by:
\beq
\mtopa=\bmtopa\pm\delta\mtopa\,,
\label{mtopa}
\eeq
where the index $\alpha$ identifies unambiguously a Mellin moment
of a given observable (so that, for example, when considering all
of the five observables of table~\ref{tab:obs} and their first three
moments, as has been done in the rightmost column of 
table~\ref{tab:nlopsms-all}, $\alpha$ can take fifteen different values).
Note that this notation does {\em not} have the same meaning of the
very similar one used in sect.~\ref{sec:eff}.
The central value of the top mass that results from the combination
of the values in eq.~(\ref{mtopa}) and its standard deviation 
are taken to be:
\beqn
&&\mtop = \sum_{\alpha=1}^{M} w_\alpha \bmtopa\,,
\label{asomega}
\\
&& \sigma^2(\mtop) = \sum_{\alpha,\beta=1}^{M} 
w_\alpha V_{\alpha\beta} w_\beta\,,
\label{sigomega}
\eeqn
where the weights $w$ and the covariance matrix $V$ are defined
as follows:
\beqn
w_\alpha &=& \sum_{\beta=1}^{M} (V^{-1})_{\alpha\beta}\Big/
\sum_{\gamma,\delta=1}^{M}(V^{-1})_{\gamma\delta},
\;\;\;\; 1 \le \alpha \le M\,,\;\;\; M = {\rm dim}(V),
\label{omegadef}
\\
V_{\alpha\beta}&=&\delta_{\alpha\beta}\left(\delta\mtopa\right)^2
+\left(1-\delta_{\alpha\beta}\right)\,
\min\Big\{\left(\delta\mtopa\right)^2,
\left(\delta\mtopb\right)^2,
C_{\alpha\beta}\delta\mtopa\delta\mtopb\Big\}\,.\phantom{aa}
\label{Vours}
\eeqn
The latter definition has been adopted in keeping with what has
been done in ref.~\cite{Frederix:2010ne}, which in turn follows
closely the prescriptions of the LEP QCD Working Group~\cite{Jones:2006nq}.
The correlation matrix $C_{\alpha\beta}$, given explicitly 
below\footnote{In order to facilitate the reading of that matrix,
each row and column is labelled with the Mellin moment it corresponds
to, in the notation introduced in eq.~(\ref{mmdef}).} in
eq.~(\ref{Cmatall3}), has been computed at one given value of the 
top mass (173~GeV): we thus neglect effects possibly due to the
dependence of such correlations on the top mass, since we expect them
to be negligible, especially in the context of eq.~(\ref{Vours}).
Given that the correlation between two variables $X$ and $Y$ is defined as
\beq
C(X,Y)=\frac{\langle\left(X-\langle X\rangle\right)
\left(Y-\langle Y\rangle\right)\rangle}{\sigma_X\sigma_Y}=
\frac{\langle XY\rangle-\langle X\rangle\langle Y\rangle}
{\sigma_X\sigma_Y}\,,
\label{corrdef}
\eeq
with $\sigma_X$ and $\sigma_Y$ the standard deviations, for any two
observables $O_r$ and $O_s$ and their $i^{th}$ and $j^{th}$ moments
$\Mm{O_r}{i}$ and $\Mm{O_s}{j}$, we use eq.~(\ref{corrdef}) by
identifying $X\equiv\Mm{O_r}{i}$ and $Y\equiv\Mm{O_s}{j}$ and
proceed similarly to what is done in eq.~(\ref{mufill}); in particular,
we have:
\beq
\langle XY\rangle=\frac{1}{N}\sum_{k=1}^N
W_k\,\Big(O_r(\kin_k)\Big)^{\,i}\Big(O_s(\kin_k)\Big)^{\,j}\,.
\eeq
We also point out that the calculation of $C_{\alpha\beta}$ has been 
performed by choosing the scale of eq.~(\ref{eq:scalesMtMT}), and in the
context of an NLO+PS+MS simulations. Although, owing to the form
of eq.~(\ref{Vours}), these choices have only a moderate impact
on the central values of the combined top masses (as we have verified
by setting $C_{\alpha\beta}=0$), we emphasise again that a more refined 
procedure will lead exactly to the same conclusions: namely, the necessity 
of combining the results obtained with different observables and moments, 
and that of performing a $\chi^2$-type test on the final outcome.

In eq.~(\ref{mtopa}) the errors affecting $\bmtopa$ are symmetric.
In the case when they are asymmetric, the procedure above, and in
particular the construction of eqs.~(\ref{omegadef}) and~(\ref{Vours}),
is repeated twice, for the $+$ and $-$ errors. The two resulting
central values for the top mass need not coincide; when this happens,
the final central value is taken to be the weighted average of the two, 
with the weights defined as the inverse of the respective $\sigma^2(\mtop)$'s 
as given in eq.~(\ref{sigomega}).

\begin{landscape}
\beqn
C=
\begin{blockarray}{c@{\hspace{1pt}}ccccccccccccccc@{\hspace{1pt}}cl}
& \mymati{\Mm{1}{1}} & \mymati{\Mm{1}{2}} & \mymati{\Mm{1}{3}} 
& \mymati{\Mm{2}{1}} & \mymati{\Mm{2}{2}} & \mymati{\Mm{2}{3}} 
& \mymati{\Mm{3}{1}} & \mymati{\Mm{3}{2}} & \mymati{\Mm{3}{3}} 
& \mymati{\Mm{4}{1}} & \mymati{\Mm{4}{2}} & \mymati{\Mm{4}{3}} 
& \mymati{\Mm{5}{1}} & \mymati{\Mm{5}{2}} & \mymati{\Mm{5}{3}} & & \\
\begin{block}{(c@{\hspace{1pt}}ccccccccccccccc@{\hspace{1pt}}c)l}
& 1    & 0.91 & 0.65  
& 0.39 & 0.41 & 0.33  
& 0.58 & 0.53 & 0.39  
& 0.53 & 0.48 & 0.37  
& 0.76 & 0.70 & 0.51 & & \mymati{\Mm{1}{1}} \\
&      & 1    & 0.89  
& 0.38 & 0.45 & 0.42  
& 0.53 & 0.56 & 0.48  
& 0.48 & 0.48 & 0.42  
& 0.71 & 0.77 & 0.68 & & \mymati{\Mm{1}{2}} \\
&      &      & 1
& 0.29 & 0.40 & 0.44 
& 0.38 & 0.48 & 0.49 
& 0.34 & 0.39 & 0.38 
& 0.52 & 0.67 & 0.73 & & \mymati{\Mm{1}{3}} \\
&      &      & 
& 1    & 0.92 & 0.68 
& 0.10 & 0.12 & 0.10 
& 0.35 & 0.29 & 0.20 
& 0.52 & 0.43 & 0.28 & & \mymati{\Mm{2}{1}} \\
&      &      & 
&      & 1    & 0.90
& 0.15 & 0.16 & 0.13 
& 0.36 & 0.32 & 0.24 
& 0.54 & 0.50 & 0.38 & & \mymati{\Mm{2}{2}} \\
&      &      & 
&      &      & 1
& 0.14 & 0.16 & 0.13 
& 0.29 & 0.28 & 0.23 
& 0.45 & 0.48 & 0.42 & & \mymati{\Mm{2}{3}} \\
&      &      & 
&      &      & 
& 1    & 0.90 & 0.65
& 0.68 & 0.61 & 0.47
& 0.75 & 0.70 & 0.52 & & \mymati{\Mm{3}{1}} \\
&      &      & 
&      &      & 
&      & 1    & 0.90
& 0.63 & 0.64 & 0.57
& 0.69 & 0.75 & 0.67 & & \mymati{\Mm{3}{2}} \\
&      &      & 
&      &      & 
&      &      & 1
& 0.46 & 0.54 & 0.55
& 0.51 & 0.65 & 0.70 & & \mymati{\Mm{3}{3}} \\
&      &      & 
&      &      & 
&      &      & 
& 1    & 0.93 & 0.74
& 0.70 & 0.62 & 0.44 & & \mymati{\Mm{4}{1}} \\
&      &      & 
&      &      & 
&      &      & 
&      & 1    & 0.93
& 0.62 & 0.62 & 0.50 & & \mymati{\Mm{4}{2}} \\
&      &      & 
&      &      & 
&      &      & 
&      &      & 1
& 0.48 & 0.54 & 0.50 & & \mymati{\Mm{4}{3}} \\
&      &      & 
&      &      & 
&      &      & 
&      &      & 
& 1    & 0.92 & 0.68 & & \mymati{\Mm{5}{1}} \\
&      &      & 
&      &      & 
&      &      & 
&      &      & 
&      & 1    & 0.90 & & \mymati{\Mm{5}{2}} \\
&      &      & 
&      &      & 
&      &      & 
&      &      & 
&      &      & 1    & & \mymati{\Mm{5}{3}} \\
    \end{block}
  \end{blockarray}
\label{Cmatall3}
\eeqn
\end{landscape}

\end{document}